\documentclass[copyright,creativecommons]{eptcs}

\renewcommand{\v}[1]{\ensuremath{\mathit{#1}}}
\newcommand{\f}[1]{\textsc{#1}}
\usepackage{graphicx}
\usepackage{booktabs}
\usepackage{mathtools}
\usepackage{comment}
\usepackage{xspace}
\usepackage{wrapfig}
\usepackage{amssymb}

\newcommand{\rshift}{\ensuremath{\mathbin{>\!\!>}}}
\newcommand{\bitand}{\ensuremath{\mathbin{\&}}}
\newcommand{\bitxor}{\ensuremath{\texttt{ xor }}}
\usepackage{color}

\providecommand{\event}{ACL2 2013}
\title{Verified AIG Algorithms in ACL2}
\def\titlerunning{Verified AIG Algorithms in ACL2}

\author{Jared Davis \qquad\qquad Sol Swords
\institute{Centaur Technology Inc. \\
7600-C N. Capital of Texas Hwy, Suite 300\\
Austin, TX 78731}
\email{\{jared,sswords\}@centtech.com}
}

\def\authorrunning{Jared Davis and Sol Swords}

\newcommand{\stobj}{stobj\xspace}
\newcommand{\stobjs}{stobjs\xspace}

\newcommand{\satlink}{Satlink\xspace}

\newcommand{\cons}{\textsc{Cons}\xspace}
\newcommand{\haig}{Hons-AIG\xspace}
\newcommand{\haigs}{Hons-AIGs\xspace}
\newcommand{\haigand}{\textsc{AigAnd}\xspace}
\newcommand{\haigor}{\textsc{AigOr}\xspace}
\newcommand{\haigeval}{\textsc{AigEval}\xspace}
\newcommand{\hlookup}{\textsc{AigEnvLookup}\xspace}
\newcommand{\hons}{\textsc{Hons}\xspace}
\newcommand{\honsequal}{\textsc{Hons-Equal}\xspace}
\newcommand{\henv}{\ensuremath{\mathit{env}}\xspace}
\newcommand{\haigt}{\texttt{t}\xspace}
\newcommand{\haignil}{\texttt{nil}\xspace}
\newcommand{\haigequiv}{\textsc{AigEquiv}\xspace}

\newcommand{\nAnd}{\textsc{and}\xspace}
\newcommand{\nNot}{\textsc{not}\xspace}

\newcommand{\aignet}{Aignet\xspace}
\newcommand{\aignets}{Aignets\xspace}

\begin{document}
\maketitle

\begin{abstract}

% State the problem.
% Say why it's an interesting problem.

  And-Inverter Graphs (AIGs) are a popular way to represent Boolean functions
  (like circuits).  AIG simplification algorithms can dramatically reduce an
  AIG, and play an important role in modern hardware verification tools like
  equivalence checkers.  In practice, these tricky algorithms are implemented
  with optimized C or C++ routines with no guarantee of correctness.
  Meanwhile, many interactive theorem provers can now employ SAT or SMT solvers
  to automatically solve finite goals, but no theorem prover makes use of these
  advanced, AIG-based approaches.

% Describe what your solution achieves.

  We have developed two ways to represent AIGs within the ACL2 theorem prover.
  One representation, \haigs, is especially convenient to use and reason
  about.  The other, \aignet, is the opposite; it is styled after modern AIG
  packages and allows for efficient algorithms.  We have implemented functions
  for converting between these representations, random vector simulation,
  conversion to CNF, etc., and developed reasoning strategies for verifying
  these algorithms.

% Say what follows from your solution.

  Aside from these contributions towards verifying AIG algorithms, this work
  has an immediate, practical benefit for ACL2 users who are using GL to
  bit-blast finite ACL2 theorems: they can now optionally trust an
  off-the-shelf SAT solver to carry out the proof, instead of using the
  built-in BDD package.  Looking to the future, it is a first step toward
  implementing verified AIG simplification algorithms that might further
  improve GL performance.

\end{abstract}

\section{Introduction}
\label{sec:introduction}

% Describe the problem.

Fully automatic tools like SAT and SMT solvers are now commonly used by
interactive theorem provers to carry out certain proofs.  For instance, in
previous work~\cite{11-swords-gl} we developed a way to solve finite ACL2
problems using a BDD package or a SAT solver.  Similarly,
Fox~\cite{11-fox-blasting} implemented a translation from bit-vector problems
in HOL4 into SAT problems which---thanks to integration work by Weber and
Amjad~\cite{09-weber-sat}---can be solved by zChaff or MiniSat.  Going further,
B\"{o}hme, et al.~\cite{11-bohme-smt} present a method to solve bit-vector
problems from Isabelle/HOL and HOL4 with the Z3 SMT solver.

% BOZO probably want to add Armand et. al's CPP paper if we can get a copy of
% it.  It sounds like they've done something for a subset of Coq now.  But
% inria's pages seem down right now.

State of the art SAT and SMT solvers are heavily optimized C or C++ programs
that are not intrinsically trustworthy.  For instance, Brummayer and
Biere~\cite{09-brummayer-fuzzing} report bugs in many SMT solvers that lead to
crashes or---much worse---wrong answers!  For SAT solvers we have some
options for dealing with this:
\begin{itemize}

\item We could simply trust the solver.  This is pragmatic and allows us to use
  the fastest tools with minimal overhead, at the risk of getting a wrong
  answer.

\item We could check the solver's work.  In the HOL family of theorem provers
  this checking is done in the usual LCF~\cite{79-milner-lcf} style, which is
  quite satisfying but is also expensive.  Faster checking is possible for
  provers that support reflection, e.g., Darbari, et al.~\cite{10-darbari-sat}
  have implemented and reflectively verified a checker for SAT proofs in Coq.
  Some important SAT techniques cannot be checked with these methods, but
  Wetzler et al.~\cite{13-wetzler-sat} describe promising work to address this.

\item We could verify the solver itself.
  Mari\'{c}~\cite{09-maric-verified-sat} has formally verified many algorithms
  used in SAT solvers.  Oe, et al.~\cite{12-oe-versat} have developed and
  verified a solver in Guru, whose implementation involves efficient, low-level
  structures like pointers, mutable arrays, and machine arithmetic.  These
  efforts are inspiring, although performance is still not competitive with
  modern solvers.

\end{itemize}
For SMT solvers there are similar approaches.  For instance, the HOL connection
to Z3's bit-vector reasoning checks the solver's work in the LCF style, and
Armand et al.~\cite{11-armand-smt-coq} have developed a reflectively verified
Coq checker for certain SMT theories.

This is a good start, but many hardware verification techniques go well beyond
SAT.  For instance, in Brayton and Mishchenko's~\cite{10-brayton-abc} ABC
system, circuits and specifications are typically represented as And-Inverter
Graphs (AIGs).  Algorithms like AIG rewriting~\cite{06-mishchenko-rewriting}
can significantly reduce the size of these graphs by analyzing their structure.
Algorithms like fraiging~\cite{05-mischenko-fraigs} can simplify AIGs by using
a combination of random simulation and SAT to find nodes that are equivalent
and can be merged.  These, and many other AIG algorithms, can greatly improve
equivalence checking.

Much like SMT solvers, these algorithms are implemented as optimized, tricky
C++ routines that might easily have errors.  We are not aware of any work to
formally verify these tools.  Although ways have been
proposed~\cite{07-chatterjee-resolution} to emit proofs from some of these
algorithms, existing tools do not generate such proofs.  This
leaves us with the choice of either (1) pragmatically using these tools, while
accepting that they may not be correct, or (2) conservatively rejecting these
tools as too risky, forgoing their benefits.

% BOZO unless that's what Nathan, et al. are trying to do.  But maybe it's
% still fair?

% BOZO add something about reachability or PDR above?

This paper presents some results toward verifying AIG algorithms and
incorporating these algorithms into ACL2.  We have actually developed two
complementary AIG representations:

\begin{itemize}

\item A Hons-based representation that is especially simple and easy to reason
  about.  The ``graph'' part of the And-Inverter Graph is kept outside of the
  logic by using the hash-consing features of ACL2(h), which avoids the sort of
  invariant-preservation theorems that you normally need when dealing with
  imperative structures.  During proofs, we abstract away the details of the
  \haig representation and instead focus on semantic equivalence, a relation
  that is amenable to Greve-like~\cite{09-greve-quantifiers} quantifier
  automation.  We have implemented and verified algorithms on \haigs such as
  random vector simulation and conversion to BDDs. (Section
  \ref{sec:hons-aigs})

\item A \stobj~\cite{02-boyer-stobjs} based representation that is styled after
  ABC's and aimed squarely at efficiency.  Here, the graph is an explicit
  \stobj array.  This is much harder to reason about: the graph is
  destructively updated as nodes are added, so we need invariant- and
  semantics-preservation theorems.  To deal with this, we develop an
  interesting \emph{extension} relationship between AIG networks, and a
  \texttt{bind-free}-based strategy for using this relation.  We have
  implemented and verified algorithms like random vector simulation, and also
  an efficient, somewhat ``smart'' conversion to Conjunctive Normal Form (CNF),
  which allows us to export AIGs to SAT. (Section \ref{sec:aignet})

\end{itemize}
Besides these contributions to reasoning about AIG algorithms, we have
integrated our \stobj based representation and its CNF conversion algorithm
into the GL~\cite{11-swords-gl} framework for bit-blasting finite ACL2
theorems.  This allows GL to solve more theorems automatically, and opens up
interesting possibilities for further scaling. (Section \ref{sec:gl}).

\section{A Hons-Based AIGs Representation}
\label{sec:hons-aigs}

ACL2(h) is an extension of ACL2 that was originally developed by Boyer and
Hunt~\cite{06-boyer-acl2h}.  It extends ACL2 with hash-consing and memoization
features that make it quite easy to develop a Hons-based AIG library that is
flexible, quite easy to prove theorems about, and reasonably efficient.

\subsection{Representation and Semantics}

(Note: our \haig representation and semantics are briefly described in previous
work by Swords and Hunt~\cite{10-swords-bddify}, we repeat some details here to
make this paper more self-contained.)

In the \haig library, an AIG is an object that represents a single Boolean
function as a tree of \nAnd and \nNot nodes with constants and input variables
as the leaves.  We use \haigt and \haignil to represent the constants true and
false, and treat any other atom as a variable.  We encode \nNot nodes as conses
of the form \texttt{($a$~.~\haignil)}, and \nAnd nodes with any other cons,
\texttt{($a$~.~$b$)}.  This special-casing of \haignil may seem awkward, but it
lets us represent either kind of node with just one cons and, at any rate, it
wouldn't be useful to \nAnd together \haignil with another AIG anyway.

The semantics of a \haig are given by an evaluation function, \haigeval, that
uses an \emph{environment} to give values to the variables:
\[
 \haigeval(x,\henv) \triangleq \begin{dcases*}
    \haignil & when $x = \haignil$, \\
    \haigt   & when $x = \haigt$, \\
    \hlookup(x,\henv) & when $x$ is an atom, \\
    \neg \haigeval(a, \henv) & when $x = \texttt{($a$~.~nil)}\textrm{, or}$ \\
    \haigeval(a,\henv) \wedge \haigeval(b,\henv) & where $x = \texttt{($a$~.~$b$)}.$ \\
 \end{dcases*}
\]

Is it fair to call these AIGs?  Implicit in the very name ``And-Inverter
\emph{Graph}'' is the idea of using a directed, acyclic graph (DAG) to
represent a collection of Boolean functions.  Where's the DAG here?  We
construct \haigs using \hons, the hash-consing constructor from ACL2(h).
Logically, \hons is nothing more than \cons.  But in the execution,
$\hons(a,b)$ checks whether there is already a hons pair corresponding to
\texttt{($a$~.~$b$)} and, if so, returns the existing pair instead of making a
new one.  The bookkeeping that keeps track of honses is done in ACL2(h) via
Common Lisp hash tables.  These tables are extended when we build new \haigs,
and, in some sense, contain the DAG for all \haigs.

We usually build \haigs with constructors named \haigand, \haigor, etc., that
apply basic simplifications like constant folding and reduce expressions like
$a \wedge a$ and $a \wedge \neg a$.  Identifying expressions like $a \wedge a$
would be very expensive if we were to use a deep structural-equality check, but
ACL2(h) has a \honsequal function that boils down to pointer-equality for
honses.  Correctness theorems for these constructors are trivial and are
stated in terms of \haigeval, e.g., for \haigand we prove
\[
\haigeval(\haigand(a,b),\henv) = \haigeval(a,\henv) \wedge \haigeval(b,\henv).
\]

\subsection{AIG Traversal and Memoization}
\begin{wrapfigure}[10]{r}[0pt]{1.2in}
\vspace{-1.7em}
%\hrule
\includegraphics[width=1.15in]{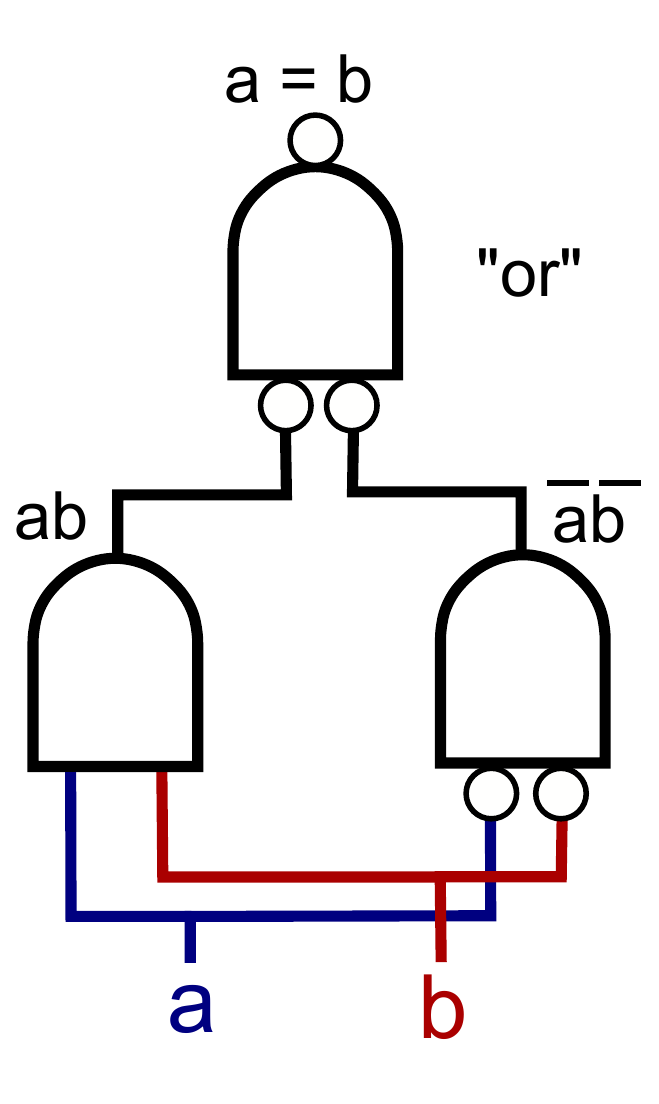}
%\hrule
\end{wrapfigure}

Almost any algorithm that operates on \haigs needs a way to avoid repeatedly
visiting the same node.  For instance, to the right is an AIG that represents
$a = b$.  Notice in particular how $a$ and $b$ are reused as
\emph{fanins}---inputs to \nAnd gates---shared by both $ab$ and $\overline{ab}$.
Because of this sharing, if we evaluate this AIG with \haigeval as above, we
will end up evaluating $a$ \textbf{twice}: once because it is a fanin of $ab$,
and once because it is a fanin of $\overline{ab}$.  We will evaluate $b$ twice
for the same reason.  As we generate larger AIGs with lots of shared structure,
the cost of this recomputation grows exponentially.

The ACL2(h) system has a function memoization capability that we can use to
avoid this recomputation.  This mechanism is especially convenient: we simply
say, ``memoize \haigeval,'' and we are done.  The logical definition of
\haigeval is unchanged, but its executable definition is extended with a Common
Lisp hash table that records its return values: when \haigeval reaches a node
whose value has already been computed, this cached return value is returned,
making the computation linear in time and space in the number of unique AIG
nodes.

Automatic memoization works well for \haigeval, but is not suitable for some
other AIG algorithms.  Notice that as \haigeval recurs, its \henv argument
remains fixed.  In many other algorithms, there are additional parameters that
change as we recur over the AIG.  These changes ruin simple memoization,
because, e.g., when we encounter $a$ for the second time, these arguments may
now be different.  In these cases, we generally represent the memo table
explicitly as a ``fast'' association list.  This is roughly as efficient as
automatic memoization; fast alists provide hash-table speeds for honsed keys.
Unfortunately, it is far less convenient.  When we verify these kind of
algorithms, we typically need invariants that say the memo tables have correct
entries.

% BOZO probably add an explicit example of this sort of algorithm.  What would
% be simplest?  Aig-vars-fast?

\subsection{Reasoning about \haigs}

\haigs are a \emph{non-canonical} representation, meaning that there are
different ways to represent the same Boolean function.  For instance, we could
represent the function $\texttt{a} \wedge \texttt{b}$ as \texttt{(a~.~b)}, or
as \texttt{(b~.~a)}, or \texttt{(a~.~(b~.~t))}, and so on.  Because of this,
when we reason about algorithms that manipulate AIGs, we typically do not want
to deal with structural equality, but rather semantic equivalence,
\[
\haigequiv(a,b) \triangleq \forall \henv : \haigeval(a,\henv) = \haigeval(b,\henv).
\]

This equivalence relation has a strong analogue in set theory.  Instead of
thinking about an AIG as some particular cons tree, think about it as a set
whose elements are the environments that satisfy it.  That is, for the AIG $a$,
think of the set $\{ \henv : \haigeval(a, \henv) = \haigt \}$.  Seen this way,
$\haigequiv(a,b)$ is nothing more than set equality; \haigand is set
intersection, \haigor is union, etc.

To reason about \haigequiv in ACL2, we use the \emph{witness}\footnote{See
  \texttt{clause-processors/witness-cp} in the ACL2 Books.} clause processor,
which is much like Greve's \texttt{quant} system~\cite{09-greve-quantifiers}
for automatically Skolemizing and instantiating quantified formulas.  When ACL2
tries to prove a conclusion of the form $\haigequiv(a,b)$, this clause
processor reduces the problem to showing that $a$ and $b$ have the same
evaluation under a particular $\henv$, similar to the pick-a-point strategy
used in Davis' osets library~\cite{04-davis-osets}.  Going further, when
we have a hypothesis of the form $\haigequiv(a,b)$, the clause processor will
add a witnessing environments $\henv$ for which $a$ and $b$ are known to
produce the same evaluation.

% BOZO I wish I could explain better why this is useful, but when I try it gets
% too long.  Maybe just refer people to Greve's paper?

%\subsection{Verified Algorithms and Applications}

We have implemented and verified several algorithms on \haigs.  As basic
building blocks, we have routines for collecting the variables in an AIG and
composing AIGs with other AIGs.  We have also developed routines for random
vector simulation; this is like \haigeval, except that we do $n$ simulations at
a time using $n$-bit integers by replacing Boolean AND and NOT with bitwise AND
and NOT.  We also developed a basic way to partition AIGs into equivalence
classes by first using random vector simulation as a coarse filter, then use
SAT to differentiate the AIGs that were the same across random simulations.
The most sophisticated algorithm that has been verified atop \haigs is the AIG
to BDD conversion algorithm of Swords and Hunt~\cite{10-swords-bddify}.  This
algorithm avoids converting irrelevant parts of the AIG, which is important
since constructing BDDs is often prohibitively expensive.

\subsection{Critique of \haigs}

\haigs have a lot of redeeming qualities.  By hiding the details of their DAG
representation beneath ACL2(h)'s implementation of hash consing, we can think
of AIG ``nodes'' as if they are ordinary ACL2 objects.  An important
consequence is that we can embed \haigs within other ACL2 objects, e.g., we
often deal with lists or alists of \haigs.  Hiding the graph from the logic
means we don't need to prove any invariant-preservation theorems, and even the
memoization needed to avoid repeating computations can often be hidden from the
logic, thanks to ACL2(h)'s memoize command.

Unfortunately, \haigs are not especially efficient.  Conventional AIG
representations use topologically ordered arrays of nodes.  By comparison, the
nodes of \haigs, being conses, have worse memory locality and must be traversed
during garbage collection, reducing overall performance.

Algorithms that operate over \haigs have to do a lot of hashing.  Even for
something as simple as \haigeval, for instance, we look up each \nAnd node in a
memoization table.  This means hashing on the address of a cons regardless of
whether we use ACL2(h) memoization or fast-alists for the table.  In contrast,
in an array-based representation, each node has a successive index and so,
e.g., \haigeval could simply allocate an array and use array-indexing, instead
of hashing, to do the memoization.  In packages like ABC, some node
representations include scratch space that can be used for such purposes,
giving even better memory locality.  Our memoization schemes also require
allocation, hash-table growth, etc., and ultimately impact garbage collection
performance.

\haigs also have one node for every \nAnd and \nNot operation, whereas most AIG
representations avoid \nNot nodes by encoding negation information in the
fanins of \nAnd nodes.  For rough perspective, we measured the \haig for a GL
proof that shows our media unit correctly implements the x86 PSADBW
instruction.  We found 53.3\% \nAnd nodes and 46.7\% \nNot nodes.  In short,
the \haig representation uses about twice as many nodes as a typical
representation uses.  This is unfortunate for memory locality and makes AIG
construction more expensive since we have to hash to create \nNot nodes.

% We aren't that space inefficient.  We don't provide nice space for scratchwork,
% which is probably good for space locality.  The ability to drop the strashing
% table isn't super critical: we can do the same thing by throwing away the hons
% space, I suppose.

% We are also not sequential.  Maybe this is explainable as: we don't need to be,
% because we can have aig-alists.  We need AIGNET to be sequential to because...?

\section{\aignet Library}
\label{sec:aignet}

In contrast to the \haig library, our \aignet library is designed for better
execution efficiency at the expense of logical simplicity. Its implementation
is similar to those in high-performance AIG packages such as ABC
\cite{10-brayton-abc}.

\subsection{Representation}

In \aignet, an AIG is represented in a single-threaded object
(\stobj)~\cite{02-boyer-stobjs}.  The \aignet \stobj may be thought of as a
medium for storing a network of inter-related (combinational) Boolean formulas.
Unlike \haigs, it also directly supports viewing these formulas as a
(sequential) finite state machine.

The primary content of the \aignet \stobj is an array of nodes.  Each node has
an \textit{ID} which is its index in the array.  We usually refer to stored
formulas via a \textit{literal}, which is a combination of an ID with a bit
saying whether it is negated or not.  A literal is represented as a natural
number using the construction
\[ \f{MkLit}(\v{id}, \v{neg}) \triangleq 2\ \v{id} + \v{neg}. \]
Extracting the parts of a literal can be implemented efficiently; in C-like notation,
\begin{align*}
\f{LitId}(\v{lit}) &\triangleq \v{lit} \rshift 1 \\
\f{LitNeg}(\v{lit}) &\triangleq \v{lit} \bitand 1.
\end{align*}
Every \aignet has a constant-false node with ID 0.  Hence, the literal 0
means constant-false, and the literal 1 means constant-true.

Here is an example circuit and its node array representation:

%\begin{figure}[h]
\begin{minipage}{0.5\textwidth}
%\begin{center}
\includegraphics[height=2.6in]{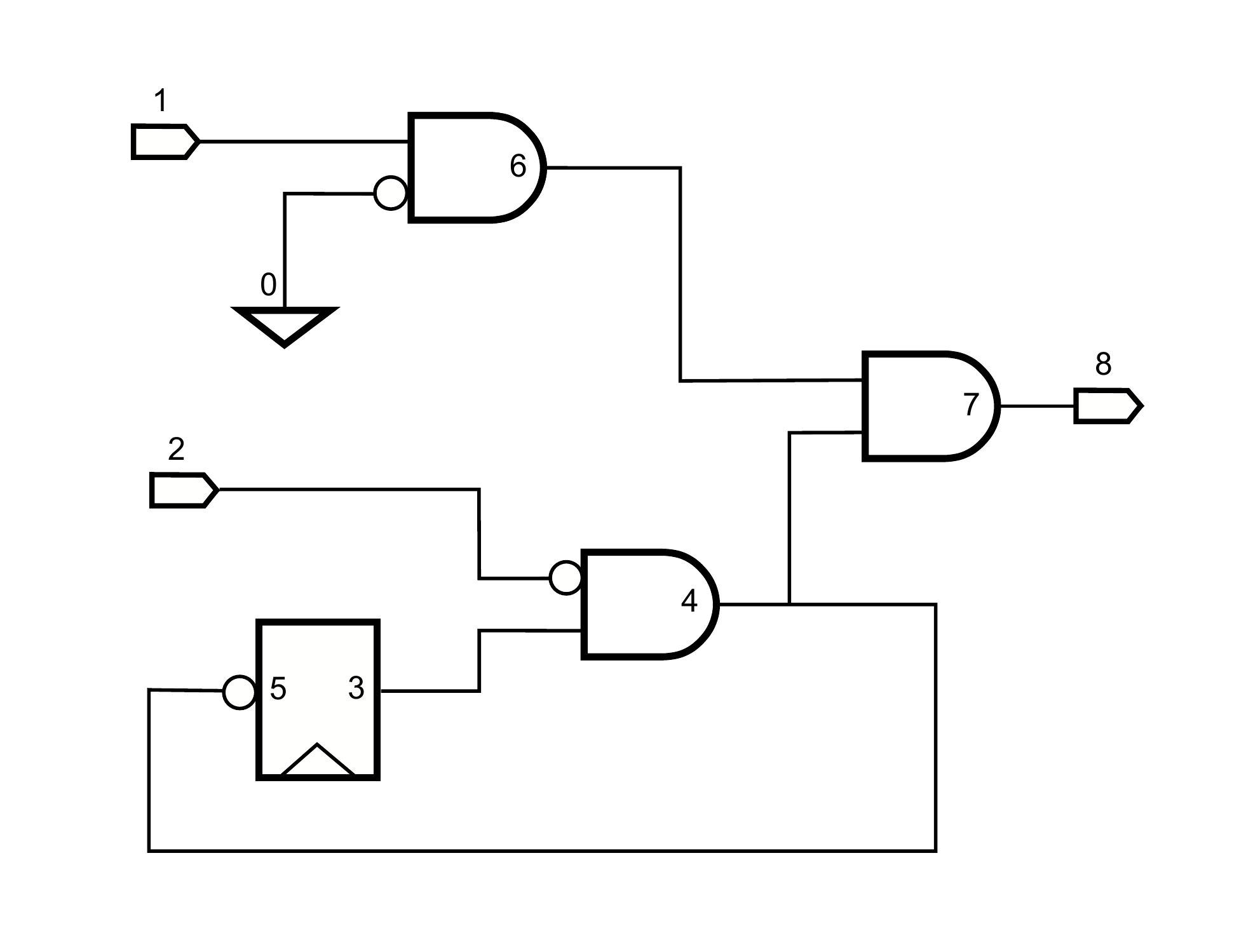}
%\end{center}
\end{minipage} \qquad
\begin{minipage}{0.5\textwidth}
%\begin{center}
\begin{tabular}{@{}lccr@{}}
\toprule
ID & Node Type & Fanins & Other \\
\midrule
0 & Constant False & & \\
1 & Primary Input & & \\
% PI \#0
2 & Primary Input & & \\
% PI \#1
3 & Register Output & & \\
%  Reg \#0
4 & AND gate & 5, 6 & \\
5 & Register Input & 9 & RO: 3 \\
6 & AND gate & 1, 2 & \\
7 & AND gate & 8, 12 & \\
8 & Primary Output & 14 & \\
%  PO \#0
\bottomrule
\end{tabular}
%\end{center} 
\end{minipage}
%\caption{Example Circuit}
%\label{aignet_fig}
%\end{figure}

Each node in the array has a type.  This circuit contains nodes of all six
types: constant false, primary input (PI), primary output (PO), \nAnd gate,
register input (RI), and register output (RO).  Unlike \haigs, there are no \nNot
gates in the node array.  Instead, the fanins of each node are encoded as
literals, which may or may not be negated.  Negations are shown as bubbles in
the picture.  For instance: the fanin of node 5 is the negation of node 4,
so its fanin literal is $2 \cdot 4 + 1 = 9$; the fanin of node 8 is node 7,
without negation, so its fanin literal is 14.

Register input and output nodes allow an \aignet to represent a sequential
circuit or finite state machine.  However, in many algorithms, the circuit is
viewed as purely combinational.  In these cases, register outputs are treated
similarly to primary inputs, both being essentially free variables; register
inputs are treated like to primary outputs, both being essentially a way to
label certain formulas.  Thus, we sometimes refer to primary inputs and
register outputs collectively as \textit{combinational inputs} (CIs), and
primary outputs and register inputs collectively as \textit{combinational
  outputs} (COs).

%It is not always convenient to refer to formulas within a network simply by
%node ID. 

% Idea:

% Primary inputs are easy to understand: they are like the variables in a \haig.
% But what are primary outputs?  When we have a collection of important \haig
% formulas, we tend to keep them in a list; any transformations or analyses are
% applied to each element of the list, and we collect up a list of answers
% corresponding to the orignal list of \haigs.  We could do a similar thing with
% a collection of aignet literals.  However, a problem we can run into is that
% node IDs (and therefore literals) are not very reliable when AIGs are being
% simplified.  For example, we might simplify a network in a manner that
% preserves the meanings of formulas referenced by the combinational outputs but
% not ``internal'' nodes.  That is, a node in the new network may have no
% relation to whatever node has the same ID in the old network.

Primary inputs are easy to understand: they are like the variables in a \haig.
But what are primary outputs?  Any AIG simplification tends to affect the
numbering of nodes: nodes may be reordered, eliminated, etc.  But by
convention, we preserve the ordering among primary outputs.  The sole purpose
of primary outputs is to maintain an order, so that before and after a
transformation, the $n$th primary output always refers to the same formula.
For similar reasons, we also generally expect that the order of primary inputs
and (for combinational simplification, at least) register outputs will remain
the same.

The \aignet \stobj supports this convention by maintaining arrays that allow us
to look up the ID of, e.g., the $n$th primary input, and also to do reverse
lookups.  These extra mappings are not material to the logical meaning of the
\stobj, which is completely expressed by the node table; e.g., looking up the
$n$th input in the input array is always the same as scanning the node table
for the $n$th PI node.

%  reorders the nodes so that nodes of the same ID in the two
% networks do not correspond.  A useful convention in these cases is to ignore
% the ordering and number of the nodes, except so as to preserve the ordering
% among the primary inputs, primary outputs, and possibly registers.  Then, one
% can refer to, say, ``the Nth primary output'' in both networks and have
% analogous nodes, even though they generally have different IDs.

% Our semantics for evaluating AIGs in aignet read their inputs

% Similarly, we preserve the orderings among PI nodes across transformations, and
% among RO nodes across transformations that are intended to preserve
% combinational (rather than sequential) equivalence.

\subsection{Semantics}

Since the \aignet \stobj can be viewed as both a medium for storing
(combinational) Boolean formulas and (sequential) finite state
machines, we have both combinational and sequential evaluation
semantics; we describe combinational evaluation first, since
sequential evaluation builds on it.

\subsubsection{Combinational Semantics}
An \aignet node or literal is evaluated under an assignment of bits to the
combinational inputs.  This assignment is given by a pair of arrays
$V_{\v{in}}, V_{\v{reg}}$, where, e.g., $V_{\v{in}}[n]$ is the value to use for
the $n$th primary input, and $V_{\v{reg}}[n]$ is the value for the $n$th
register output.  Evaluation is defined as follows, except that we hide the
stobj argument for simpler presentation:
\begin{align*}
  \f{EvalLit}(\v{lit}, V_{\v{in}}, V_{\v{reg}}) &\triangleq
  \f{EvalID}(\f{LitId}(\v{lit}), V_{\v{in}}, V_{\v{reg}}) \bitxor{}
  \f{LitNeg}(\v{lit})
\\
  \f{EvalID}(\v{id}, V_{\v{in}}, V_{\v{reg}}) &\triangleq
  \begin{dcases*}
    0 & if $\f{Type}(\v{id}) = \f{Const}$ \\
    V_{\v{in}}[\f{InIdx}(\v{id})] & if $\f{Type}(\v{id}) = \f{PI}$ \\
    V_{\v{reg}}[\f{RegIdx}(\v{id})] & if $\f{Type}(\v{id}) = \f{RO}$ \\
    \f{EvalLit}(\f{Fanin1}(\v{id}), V_{\v{in}}, V_{\v{reg}}) \\
    \quad \bitand\ \f{EvalLit}(\f{Fanin2}(\v{id}), V_{\v{in}},
    V_{\v{reg}})
    & if $\f{Type}(\v{id}) = \f{And}$ \\
    \f{EvalLit}(\f{Fanin}(\v{id}), V_{\v{in}}, V_{\v{reg}}) & if
    $\f{Type}(\v{id}) \in \{\f{PO}, \f{RI}\}$
  \end{dcases*}
\end{align*}

This evaluation function is nice for reasoning but not practical for execution
since it does no memoization.  A more practical evaluation function simply does
a linear sweep over the node array, storing each node's value in a result
array.  We require the node array to be topologically ordered, so that each
fanin's value is already computed.  This makes evaluation linear in the size of
the graph.  (Topological ordering is also important to show that \f{EvalLit}
terminates!)

% We can then prove that after this function runs,
% $V_{\v{res}}[\v{id}] = \f{EvalId}(\v{id},V_{\v{in}},V_{\v{reg}})$.

% The nodes must be stored in the array in topological order with respect to the
% fanins; that is, a node's fanins must have IDs less than the node's own ID.
% BOZO  a register input node stores the ID of its corresponding
% register output, which must also be less than its own ID.

\begin{wrapfigure}[8]{r}[0pt]{2.8in}
%\begin{table}[h]
\vspace{-.5em}
\begin{center}
\begin{tabular}{@{}lrr@{}}
\toprule
Representation & Time & Allocation \\
\midrule
\haig & 57.6 sec & 2.62 GB \\
\aignet & 6.85 sec & 16.2 MB \\
ABC (no opt.) & 4.03 sec & 16.0 MB \\
ABC (opt.) & 1.21 sec & 16.0 MB \\
\bottomrule
\end{tabular}
\end{center}
%\end{table}
\end{wrapfigure}
Evaluation is a good example of the performance advantage of the \aignet
representation over the \haig representation.  We benchmarked evaluating the
functions for a hardware module with 3,664 input variables, 131,605 \nAnd
nodes, and (in the \haig representation) 121,490 \nNot nodes.  We also include
a comparison with ABC, compiled with and without GCC's ``-O'' optimization
flag.  In each case, we evaluated the network under the all-false environment
1000 times.  For the \haig test, we cleared the \f{AigEval} memoization hash
table at each iteration, and used a large enough Lisp heap to avoid garbage
collection, so that GC time was not counted. For the \aignet and ABC tests, we
used a bit array to remember node values, re-allocating and clearing this array
at each iteration.

\subsubsection{Sequential Semantics}
\newcommand{\vfr}{V_{\!\v{fr}}}
Sequential evaluation depends on an initial assignment to the
registers and a series of assignments to the primary inputs.  A common
convention is to assume that each register's initial value is 0,
but the \aignet library does not enforce this convention.
Letting $V_{\v{ini}}$ be the initial state values (indexed by register
number) and $\vfr{}$ be a two-dimensional array giving the input
assignments at each frame (indexed first by frame, then by input
number), the evaluation of a node or literal at frame $k$ is given by:
\begin{align*}
  \f{SeqEvLit}(k, \v{lit}, \vfr{}, V_{\v{ini}}) &\triangleq
  \f{SeqEvID}(k, \f{LitId}(\v{lit}), \vfr{}, V_{\v{ini}}) \bitxor
  \f{LitNeg}(\v{lit})
\\
  \f{SeqEvID}(k, \v{id}, \vfr{}, V_{\v{ini}}) &\triangleq
  \begin{dcases*}
    0 & if $\f{Type}(\v{id}) = \f{Const}$ \\
    \vfr{}[k][\f{InIdx}(\v{id})] & if $\f{Type}(\v{id}) = \f{PI}$ \\
    V_{\v{ini}}[\f{RegIdx}(\v{id})] & if $\f{Type}(\v{id}) = \f{RO}$
    and $k = 0$\\
    \f{SeqEvID}(k-1, \f{RegIn}(\v{id}), \vfr{}, V_{\v{ini}}) & if $\f{Type}(\v{id}) = \f{RO}$
    and $k > 0$\\
    \f{SeqEvLit}(k, \f{Fanin1}(\v{id}), \vfr{}, V_{\v{ini}}) \\
    \quad \bitand\ \f{SeqEvLit}(k, \f{Fanin2}(\v{id}), \vfr,
    V_{\v{ini}})& if $\f{Type}(\v{id}) = \f{And}$ \\
    \f{SeqEvLit}(k, \f{Fanin}(\v{id}), \vfr{}, V_{\v{ini}}) & if
    $\f{Type}(\v{id}) \in \{\f{PO}, \f{RI}\}.$
  \end{dcases*} \\
\end{align*}

Alternatively, we can define $\f{SeqEvID}$ in terms of the combinational semantics
\f{EvalId} as follows:
\begin{align*}
\f{SeqEvID}(k, \v{id}, \vfr{}, V_{\v{ini}}) &\triangleq
 \f{EvalId}\left(\v{id}, \vfr{}[k], \f{RegFrame}\left(k, \vfr{},
V_{\v{ini}}\right)\right) \\
 \f{RegFrame}(k, \vfr{}, V_{\v{ini}})[n] &\triangleq
\begin{dcases*}
V_{\v{ini}}[n] & if $k = 0$ \\
\f{SeqEvId}\left(k-1, \f{RegIn}\left(\f{NthReg}\left(n\right)\right),
 \vfr{}, V_{\v{ini}}\right) & if $k > 0$. \\
\end{dcases*}
\end{align*}
Here, we capture the values on the register outputs at each frame $k$ as a bit
array, $\f{RegFrame}(k, \vfr{}, V_{\v{ini}})$.  This takes the initial values
at frame 0, and for later frames collects the register input values computed
for the previous frame.  Then the sequential evaluation of any node in frame
$k$ reduces to a combinational evaluation of the inputs and register values
for that frame.

Either definition is well suited for reasoning but performs poorly on real
examples since it does no memoization.  An efficient implementation that
linearly computes each frame in sequence by storing the current values in a bit
array can be proven to compute these values.

\subsection{Constructing \aignets}

The following basic functions are used to add nodes to an \aignet \stobj.  The
\stobj is both an input and output of all functions below, but is omitted for
brevity:
\begin{itemize}
\item $\f{AddInput}() \rightarrow \v{lit}$: Adds a new primary input node
  and returns its non-negated literal.
\item $\f{AddReg}() \rightarrow \v{lit}$: Adds a new register output
  node and returns its non-negated literal.
\item $\f{AddAnd}(\v{lit1}, \v{lit2}) \rightarrow \v{lit}$: Adds a new
  \nAnd gate of $\v{lit1}, \v{lit2}$; returns its non-negated
  literal.
\item $\f{AddOutput}(\v{lit})$: Adds a new primary output with fanin
  $\v{lit}$.
\item $\f{AddRegIn}(\v{lit}, \v{ro})$: Adds a new register input node
  with fanin $\v{lit}$ and register output ID $\v{ro}$.
\end{itemize}

% ($\f{AddOutput}$ and $\f{AddRegIn}$ do not return literals because
% combinational outputs should not themselves be fanin nodes.  Their
% fanins may be fanins of other nodes, however.)

$\f{AddAnd}$ is a very low-level way to add a gate to an \aignet \stobj.  It
unconditionally adds an \nAnd node with no simplification and with no regard to
whether such a node already exists.  Higher-level functions called \f{HashAnd},
\f{HashOr}, \f{HashXor}, and \f{HashMux} can add logic in a smarter way:
\begin{itemize}
\item They may optionally propagate constants, eliminate certain
  tautologies and contradictions, detect cases where an \nAnd is
  logically equivalent to one of its fanins, etc., to produce equivalent
  formulas with fewer nodes.  These reductions are described in Brummayer
  and Biere \cite{06-brummayer-without-blowup}.
\item They allow \textit{structural hashing} (strashing): when preparing to add
  an \nAnd node, we check to see if there is an existing \nAnd node with the
  same fanins, and if so reuse it.  Our structural hashing implementation
  currently requires a trust tag to allow \stobjs to have hash table fields;
  we believe it is sound, and perhaps it can be integrated into ACL2 in the future.
\end{itemize}

%BOZO seems like this p interrupts the flow
In \haigs, structural hashing is automatic due to our use of \hons, but in our
\stobj-based approach the strash table is necessarily visible in the logic.
Normally we would need an invariant about the strash table to know that our
construction functions return nodes with the right meanings; this would mean
proving invariant-preservation theorems for many algorithms.  We initially
tried to do this, but found it to be tedious.  We then realized that it is
reasonably fast\footnote{ This runtime check costs an overhead of 30--40\% in
  the worst case, i.e., when every strash lookup is a hit and no simplification
  is performed.  We generally expect to see fewer strash hits and to
  perform simplification, which has much greater overhead.  And, at any rate,
  \aignet construction is typically not a bottleneck.}  to check whether a
hit in the strash table actually gives the correct node.  Adding this runtime
check to our strash lookup obviates the need for an invariant altogether!
Explicit hashing also has some advantages: once we are done building a network
we may throw out the strash table and reclaim its memory while keeping the AIG
itself.

Each of these functions returns a literal that provably evaluates to the
correct function of the evaluations of the input literals, and also provably
preserves the meaning of previously existing literals.  For example, in the
case of \f{HashAnd}, letting
\[
  (\v{newlit}, \v{strash}', \v{aignet'}) = \f{HashAnd}(\v{lit}_1, \v{lit}_2, \v{strash}, \v{simp\_opts}, \v{aignet}),
\]
the new-literal correctness theorem is
\[
 \begin{array}{lll}
\f{EvalLit}(\v{newlit}, V_{in}, V_{reg}, \v{aignet'}) &=& \f{EvalLit}(\v{lit}_1, V_{in}, V_{reg}, \v{aignet}) \ \wedge \\
      & & \f{EvalLit}(\v{lit}_2, V_{in}, V_{reg}, \v{aignet}),
\end{array}
\]
and the previous-literal preservation theorem is
\[
\begin{array}{lll}
  \f{LitId}(\v{oldlit}) < \f{NumNodes}(\v{aignet})
    &\Rightarrow & \f{EvalLit}(\v{oldlit}, V_{in}, V_{reg}, \v{aignet}') = \\
    &            & \f{EvalLit}(\v{oldlit}, V_{in}, V_{reg}, \v{aignet}).
\end{array}
\]

This preservation property is an instance of a more pervasive pattern: most
functions which modify an \aignet \stobj are simply adding new nodes, without
modifying the existing ones.  This means that many properties of existing nodes
in the original network remain true in the modified network.  To take advantage
of this, we use a relation
\begin{align*}
\f{Aignet}&\f{Extension}(\v{new}, \v{old}) \triangleq \\
 & \ \f{NumNodes}(\v{old}) \le \f{NumNodes}(\v{new}) \ \wedge \\
 & \
  \forall\ \v{id} < \f{NumNodes}\left(\v{old}\right) : 
  \f{Nodes}(\v{new})[\v{id}] = \f{Nodes}(\v{old})[\v{id}],
\end{align*}
that is, all nodes that were in the old network are still the same in the new
network.

Most functions that modify an \aignet \stobj produce an output \aignet which is
an extension of the input \aignet, and there are many useful preservation
properties that follow when we know this is the case.  For example, when we
know $\f{AignetExtension}(\v{new}, \v{old})$, it follows that:
\begin{itemize}
\item an ID that is within bounds for \v{old} is still in-bounds for \v{new}
\item a list of in-bounds literals for \v{old} is still in-bounds for \v{new}
\item evaluation of an in-bounds literal in \v{old} is the same as its
  evaluation in \v{new}
\item the $N$th output in \v{old} still has the same ID in \v{new}.
\end{itemize}

Written as rewrite rules, these preservation properties generally bind \v{new}
in the LHS, but have \v{old} as a free variable.  We use a
\texttt{bind-free}~\cite{05-hunt-meta} form to instruct ACL2 how to bind
\v{old}.  In particular, if \v{new} is bound to a \stobj return value of a
function call, we will try binding \v{old} to the corresponding \stobj input of
that function call.  Otherwise, we fall back to searching the type-alist for a
known-true term of the form $\f{AignetExtension}(\v{new}, \v{old})$.  This
strategy works well in most cases, but in some cases it would be better to try
multiple different bindings for \v{old}, in case the first one fails to satisfy
one of the other hypotheses.  This capability is not yet available to
\texttt{bind-free}.

In the \aignet library sources, we have identified about 40 such preservation
properties.
% about 40 rules are defined that follow
% the pattern of proving a property of an \aignet \v{new} by showing that it
% holds of another \aignet of which \v{new} is an extension.
Furthermore, we have proven that about 25 \aignet-modifying functions return an
\aignet that is an extension of their input.  If we tried to do without the
\f{AignetExtension} property and instead prove each of the 40 preservation
properties for all 25 functions then we would need 1000 rules, whereas we now
accomplish the same thing with only 65 rules.

\subsection{CNF Generation}
\label{sec:aignet-cnfgen}

We have implemented and proven correct a CNF generation algorithm for \aignet
\stobjs.  Today, this algorithm allows us to export AIG-derived problems to an
external SAT solver.  For the future, it is also groundwork toward implementing
and verifying many AIG algorithms that are based on \emph{incremental
  SAT}~\cite{03-een-extensible}, a technique for efficiently checking multiple
related satisfiability queries.  For instance, modern model- and
equivalence-checking packages make use of:

%   Some of the best algorithms in industrial-strength
% SAT-based model- and equivalence-checking packages involve checking
% satisfiability of many overlapping formulas.  These formulas are often
% derived from a single circuit by applying different sets of constraints.
% A few examples:
\begin{itemize}
\item \textit{Fraiging} or \textit{SAT sweeping}~\cite{05-mischenko-fraigs}
  is a combinational simplification algorithm designed to eliminate
  circuit nodes that are redundant; that is, any nodes whose
  functionality is provably the same as another.  It finds candidate
  combinational equivalences among circuit nodes using random
  simulation, then attempts to prove or disprove each equivalence
  using SAT.
\item \textit{Signal correlation} \cite{08-mischenko-scalable} is a sequential
  simplification algorithm that somewhat resembles fraiging, but allows
  elimination of nodes that are sequentially equivalent, even if they are not
  combinationally equivalent.  It finds candidate sequential equivalences using
  random simulation, then uses SAT to attempt to prove all equivalences by
  induction over one or more frames.

\item \textit{IC3} or \textit{property directed
    reachability}~\cite{12-bradley-ic3,11-een-pdr} is an algorithm for
  checking safety properties.  It operates by repeatedly using SAT to refine an
  overapproximation of the reachable state space until it is shown that no bad
  states are reachable or a counterexample is found.

% This algorithm
%   maintains over-approximate expressions of the reachable state space
%   at a series of time steps.  Its basic operation is to check using
%   SAT whether there are states allowed by the overapproximation that
%   can transition to a bad state (from which a violation of the
%   property is known to be reachable) in one step.  It uses the results
%   of these SAT checks to narrow down its over-approximate reachability
%   formulas until it either finds a counterexample or it can be shown
%   that no bad states are reachable.
\end{itemize}
An incremental SAT solver
maintains a database of clauses, of which some are given and some are learned.
Typically the CNF database as a whole---the conjunction of all the clauses---is
known to be satisfiable.  The problems to solve are given by providing a {\em cube} (a
conjunction of literals) as an additional constraint.  The solver checks the
satisfiability of the cube together with the clause database, but only learns
clauses relative to the database, not assuming the cube.  The point is:
subsequent SAT checks may reuse heuristic information and learned clauses even though
the constraints change.

An incremental SAT interface is useful for solving problems generated by
AIG-based algorithms.  The idea here is to generate the clause database from
the circuit by adding structural constraints as in the
Tseitin~\cite{68-tseitin-propositional} transform.  For instance, if $c$ is an
\nAnd node with fanins $a$ and $b$, we add the clauses for $\neg a \Rightarrow
\neg c$, $\neg b \Rightarrow \neg c$, and $(a \wedge b) \Rightarrow c$.  These
clauses fully and exactly represent the logical meaning of the \nAnd node: any
possible configuration of the \nAnd gate and its fanins satisfies the clauses,
and any satisfying assignment of the clauses is a possible configuration of the
\nAnd gate.

When we generate a clause database in this way, any evaluation of the AIG
induces a satisfying assignment of the database by assigning each literal its
value under the evaluation.  This allows us to prove facts about the network by
checking satisfiability of the database together with some additional
constraints as follows.  Suppose we generate a clause database from part of a
circuit.  Suppose we prove that the database together with a certain
constraint---say, some literal---is unsatisfiable.  This proves that this
literal may not be true under any evaluation of the circuit.  Why?  Suppose we
have an assignment to the combinational inputs of the network that makes this
literal true.  This produces a satisfying assignment of the database, and the
literal is true, so the constraint is satisfied, which we have proved
impossible.  This generalizes to any additional constraint not in the database,
most commonly a cube of literals.

Conversely, if the database contains sufficient structural constraints
relating some set of nodes to their fanins, and also transitively for
their fanins reaching back to the combinational inputs, then any
satisfying assignment of the database induces an assignment to the CIs
of the circuit under which each node in the set evaluates to its ID's
value under the satisfying assignment.  This allows us to transform a
satisfying assignment produced by a SAT solver into
an assignment to the combinational inputs that satisfies whatever
constraint we added to the clause database.

Our CNF generation algorithm is a commonly-implemented~\cite{07-een-synthesis}
variation on the standard Tseitin transform.  This variation has two
optimizations that reduce the number of variables and clauses that will be
given to the solver, which tends to speed up SAT checks.  First, it recognizes
\emph{supergates}: trees of multiple \nAnd gates, without negations, where only
the root has multiple fanouts.  For each supergate it generates $n$ binary
clauses and one clause of length $n+1$:
\begin{align*}
o = i_0 \wedge i_1 \wedge \ldots \wedge i_n
  & \qquad \longrightarrow \qquad
\begin{array}{rll}
% BOZO make this look nicer, maybe get rid of the left side?
% Jared: getting rid of the left side.
 \neg i_0 &\Rightarrow& \neg o \\
  \dots \\
  \neg i_n &\Rightarrow& \neg o \\
  i_0 \wedge \ldots \wedge i_n &\Rightarrow& o
\end{array}
\end{align*}
Second, it recognizes two-level nestings of gates representing a mux
(if-then-else) structure, and generates a special set of clauses:
\begin{align*}
o = \textsf{if } a \textsf{ then } b \textsf{ else } c
  & \qquad \longrightarrow \qquad
%   \begin{align*}
% % BOZO make this look nicer, maybe get rid of the left side?
%     i_0 \vee \neg o & & \neg i_0 \Rightarrow \neg o \\
%     \dots & & \dots \\
%     i_n \vee \neg o & & \neg i_n \Rightarrow \neg o \\
%     \neg i_0 \vee \ldots \vee \neg i_n \vee o & & i_0 \wedge \ldots \wedge i_n \Rightarrow o \\
%   \end{align*}
  \begin{array}{rll}
 a \wedge b &\Rightarrow& o \\
 \neg a \wedge c &\Rightarrow& o \\
 a \wedge \neg b &\Rightarrow& \neg o \\
 \neg a \wedge \neg c &\Rightarrow& \neg o \\
 b \wedge c &\Rightarrow& o \\
 \neg b \wedge \neg c &\Rightarrow& \neg o
  \end{array}
\end{align*}
%   \begin{align*}
% % BOZO make this look nicer
%     \neg a \vee \neg b \vee o & & a \wedge b \Rightarrow o \\
%     a \vee \neg c \vee o & & \neg a \wedge c \Rightarrow o \\
%     \neg a \vee b \vee \neg o & & a \wedge \neg b \Rightarrow \neg o \\
%     a \vee c \vee \neg o & & \neg a \wedge \neg c \Rightarrow \neg o \\
%     \neg b \vee \neg c \vee o && b \wedge c \Rightarrow o \\
%     b \vee c \vee \neg o && \neg b \wedge \neg c \Rightarrow \neg o \\
%   \end{align*}
The last two clauses are unnecessary but are included to improve SAT
performance~\cite{07-een-synthesis}; they are omitted in the degenerate case
where $b=\neg c$, i.e. an XOR gate.

The CNF generation function has the following signature:
\[ \f{AddCnf}(\v{id}, \v{marks}, \v{cnf}, \v{aignet})
\rightarrow (\v{marks}', \v{cnf}'). \]
Here, \v{cnf} is the accumulator for clauses, and \v{marks} is a bit
array that tracks the identifiers whose structural
constraints (recursively, back to the CIs) are encoded in the CNF.

We prove that \f{AddCnf} marks \v{id} and preserves an invariant,
$\f{CnfForAignet}(\v{cnf}, \v{marks}, \v{aignet})$.  When this invariant
holds, we can map evaluations of \v{aignet} to satisfying assignments
of \v{cnf} and vice versa, as described above.  To state the invariant
we need some additional definitions:
\begin{itemize}

\item $\f{CnfEval}(\v{cnf}, \v{env})$ evaluates a CNF formula under a
  bit array $env$ mapping IDs to values.

\item $\f{AignetEval}(\v{aignet}, \v{env})$ updates \v{env} so that
  the values assigned to CI IDs are preserved, but for all other IDs,
  the values under those CI assignments are stored.

\item $\f{MarkedValsCorrect}(\v{aignet}, \v{marks}, \v{env})$ is true when
  \v{env} binds every marked ID to its evaluation under the CI assignments,
  i.e., when
\[ \forall \v{id} : \v{marks}[\v{id}] \Rightarrow \f{AignetEval}(\v{aignet}, \v{env})[\v{id}] = \v{env}[\v{id}].\]

\end{itemize}
We can now define our invariant:
\begin{align*}
\f{CnfForAignet}& (\v{cnf}, \v{marks}, \v{aignet}) \triangleq \\
\forall \v{env} : & \ (\f{CnfEval}(\v{cnf}, \v{env}) \Rightarrow \f{MarkedValsCorrect}(\v{aignet}, \v{marks}, \v{env})) \\
& \wedge \f{CnfEval}(\v{cnf}, \f{AignetEval}(\v{aignet}, \v{env})).
\end{align*}
The two conjuncts are the two directions.  The first says, for any satisfying
assignment of the CNF, how to obtain a consistent evaluation of the circuit.
The second says, given an evaluation of the circuit, how to obtain a satisfying
assignment of the CNF.

When \v{cnf} is empty and no nodes are marked, \f{CnfForAignet} is trivially
true.  This is the starting point for adding nodes to \v{cnf}.  Then, we prove
our invariant is preserved by \f{AddCnf}, i.e.,
\begin{align*}
& \mathsf{let}\ (\v{marks}', \v{cnf}') = \f{AddCnf}(\v{id}, \v{marks},
\v{cnf}, \v{aignet}) \ \mathsf{in} \\
& \quad \f{CnfForAignet}(\v{cnf}, \v{marks}, \v{aignet}) \Rightarrow
\f{CnfForAignet}(\v{cnf}', \v{marks}', \v{aignet}).
\end{align*}
% BOZO this is harder than it looks
% We also prove that adding nodes to the aignet preserves the invariant:
% property:
% \begin{multline*}
%  \f{CnfForAignet}(\v{cnf}, \v{marks}, \v{aignet}) \wedge
%  \f{AignetExtension}(\v{aignet}', \v{aignet}) \Rightarrow \\
%  \f{CnfForAignet}(\v{cnf}, \v{marks}, \v{aignet}').
% \end{multline*}

When we know that \f{CnfForAignet} holds, then we know that satisfiability
proofs about the CNF (with additional constraints) imply
satisfiability results about the circuit.  For example, we define a
function $\f{AignetCisToCnfEnv}(V_{\v{in}}, V_{\v{reg}}, \v{aignet})$ and prove
\begin{align*}
& \  \f{CnfForAignet}(\v{cnf}, \v{marks}, \v{aignet}) \\
\wedge & \ \f{LitEval}(\v{lit}, V_{\v{in}}, V_{\v{reg}}, \v{aignet}) \\
\wedge  & \  \v{marks}[\f{LitId}(\v{lit})] \\
\Rightarrow \quad & \\
&  \f{CnfEval}\left(\v{cnf} \cup \left\{ \v{lit} \right\},
\f{AignetCisToCnfEnv}\left(V_{\v{in}}, V_{\v{reg}}, \v{aignet}\right)\right).
\end{align*}
Therefore if $\v{cnf} \cup \{\v{lit}\}$ is unsatisfiable, then there
is no $V_{\v{in}}, V_{\v{reg}}$ for which $\f{LitEval}(\v{lit},
V_{\v{in}}, V_{\v{reg}}, \v{aignet})$.
Conversely, we have $\f{CnfEnvToAignetCis}(\v{env}, \v{aignet})$ for
which we prove:
\begin{align*}
& \ \f{CnfForAignet}(\v{cnf}, \v{marks}, \v{aignet}) \\
\wedge & \ \f{CnfEval}\left(\v{cnf} \cup \left\{ \v{lit} \right\}, \v{env}\right) \\
\wedge & \ \v{marks}[\f{LitId}(\v{lit})] \\
\Rightarrow \quad & \\
& \mathsf{let}\ (V_{\v{in}}, V_{\v{reg}}) =
\f{CnfEnvToAignetCis}(\v{env}, \v{aignet}) \ \mathsf{in} \\
& \quad \f{LitEval}(\v{lit}, V_{\v{in}}, V_{\v{reg}}, \v{aignet}).
\end{align*}
That is, a satisfying assignment for $\v{cnf} \cup \left\{ \v{lit}
\right\}$ can be converted to a CI assignment under which \v{lit} is
satisfied.

\section{GL Integration}
\label{sec:gl}

In previous work~\cite{11-swords-gl} we introduced \emph{GL}, a framework for
automatically proving ``finite'' ACL2 theorems.  GL works by symbolically
executing ACL2 formulas to \emph{bit-blast} (translate) them into Boolean
formulas, which can then be solved using a tool like a BDD package or a SAT
solver.

We use GL to verify x86 execution units~\cite{10-hardin-centaur}.  Its ability
to solve these problems depends on the performance of its symbolic
execution and the capacity of these solvers.  Without help, GL can
verify many instructions within a few seconds, and some others in a few
minutes.  For harder instructions, we have to manually intervene, e.g., we
might decompose the proof in some way.

We want to scale up GL to avoid this human effort.
Toward this end, we have developed a new GL \emph{mode} that uses \aignet and
its CNF conversion algorithm to connect GL to any off-the-shelf SAT solver that
implements the DIMACS format.  As we will explain shortly, this already has
advantages over GL's previous approach to using SAT, and it opens up a
promising route for future improvements, viz. simplifying the AIG before
calling SAT.

GL can operate in different modes that govern how Boolean formulas will be
represented during the symbolic execution, and how these formulas will be
solved.  The mode to use can be chosen at run-time on a per-proof basis using
attachments~\cite{11-kaufmann-defattach}, and this choice can have a
significant impact on GL's performance.  Previously, GL has supported three
modes: \emph{BDD mode}, \emph{AIG mode}, and \emph{AIG-Cert mode}.  We describe
these and our new \emph{\satlink mode} modes below.  But first, here is a
sketch:

%\vspace{.1em}
\begin{center}
\includegraphics[width=.96\textwidth]{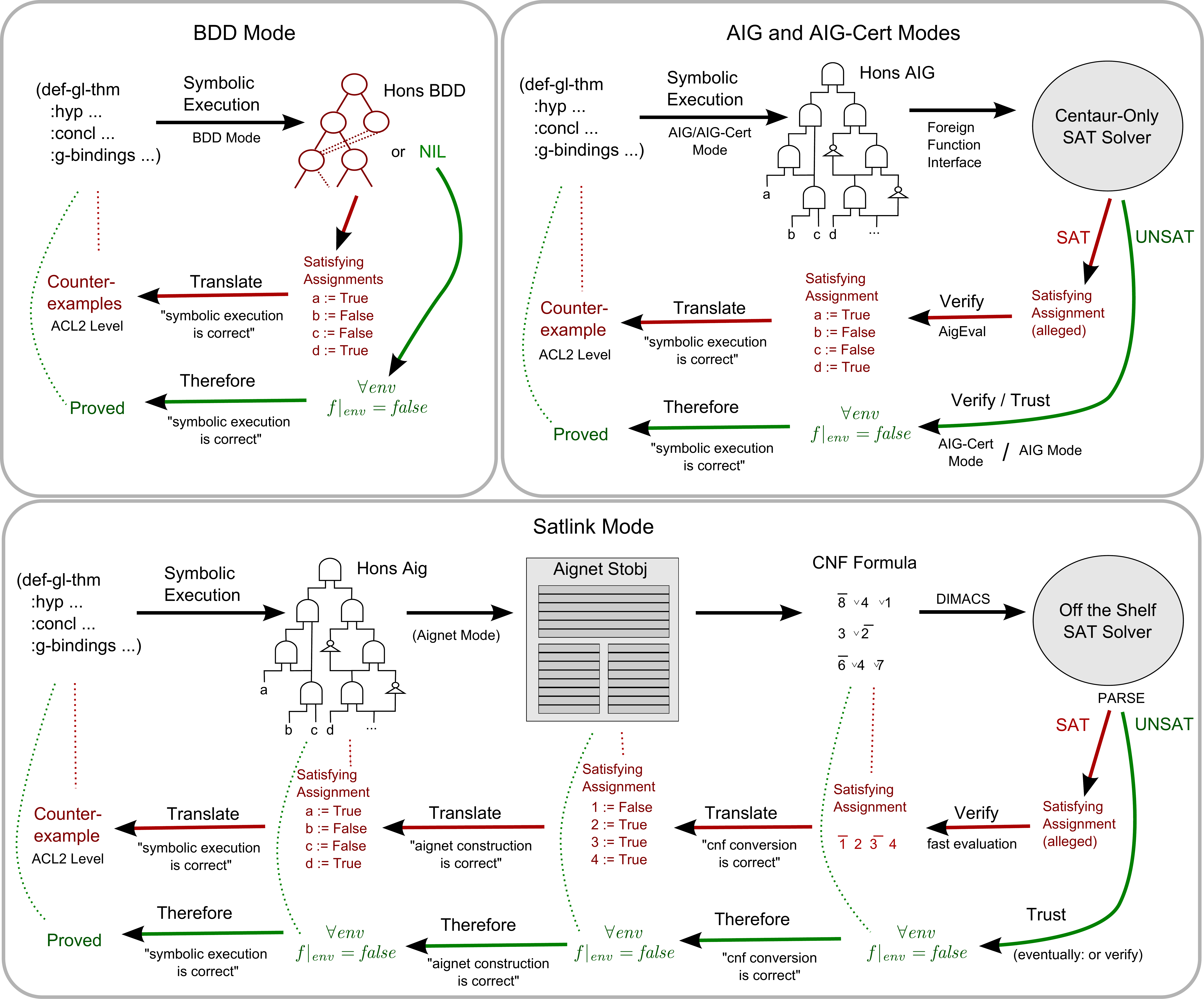}
\end{center}

In BDD mode (the default), GL uses a simple Hons-based BDD
representation~\cite{06-boyer-acl2h} of Boolean functions during symbolic
execution, and no back-end solver is needed because the truth of a BDD is
immediately evident.  This works well when the user knows how to choose a good
BDD ordering, but is not well-suited for larger, less-structured problems.

In AIG and AIG-Cert modes, \haigs are used as the Boolean function
representation during symbolic execution, and a SAT solver is used as the
back-end.  These modes are convenient since a good BDD ordering is not needed,
and they tend to outperform BDDs on less-structured problems.  In AIG mode,
the SAT solver's claims of UNSAT are trusted without verification; in
AIG-Cert mode they are checked using a verified checker developed
by Matt Kaufmann.  Both AIG modes are integrated with a SAT solver that
can directly accept AIGs as inputs, but which unfortunately we cannot release.

Our new mode, \satlink mode, still uses \haigs as the Boolean function
representation during symbolic execution.  We then convert the goal \haig into
an \aignet and use our verified CNF conversion algorithm to generate clauses
that can be sent to any SAT solver that implements the usual DIMACS format.
Like AIG mode, the SAT solver is simply trusted to be correct.  We have not yet
implemented a proof-checking scheme for this \satlink approach, but we are
optimistic that recent work by Wetzler, et al.~\cite{13-wetzler-sat} might lead
to a high-quality way to fill in this gap.

\begin{wrapfigure}[6]{r}[0pt]{3.1in}
\vspace{-1.5em}
\begin{center}
\begin{tabular}{lll}
\toprule
128-bit PSADBW    & Parameterized & Full \\
\midrule
AIG Mode          & 84.38 sec     & ? ($>$24 hr) \\
\satlink/Lingeling & 62.02 sec     & 50.08 sec \\
\satlink/Glucose   &  6.11 sec     & 14.88 sec \\
\bottomrule
\end{tabular}
\end{center}
\end{wrapfigure}

Already, \satlink mode appears to be quite promising.  For instance, we applied
it to a somewhat difficult proof: the correctness of our media unit's 128-bit
x86 PSADBW instruction.  We have never successfully proved this theorem with
BDDs.  Even for AIG mode, performance was so bad that we spent time manually
\emph{parameterizing}~\cite{11-swords-gl} the theorem.  State-of-the-art tools
like Glucose~\cite{09-audemard-glucose} and Lingeling~\cite{12-biere-lingeling}
can solve the full problem quickly, making this work unnecessary!

% There is a cost to verifying the SAT solver's answers.  One kind of overhead is
% what you would expect: some time is needed to record proofs in the SAT solver,
% and some time is needed to check them in ACL2.  But another kind of overhead is
% due to CNF conversion.  Our SAT solver accepts problems directly in AIG format.
% When it is not recording proofs, it uses a smart CNF conversion like our
% \aignet algorithm (Section \ref{sec:aignet-cnfgen}).  But when proof-recording
% is enabled, it uses a simplified CNF generation, with no supergate/mux
% optimization, so that its clauses will agree with the checker.  The short of it
% is: the back-end SAT engine gets a harder set of clauses, which can lead to a
% significant slowdown.  [BOZO insert anecdotal performance figure here.]

A promising direction for future work is to simplify the \aignet before
converting it to CNF.  To explore this, we developed an unverified, experimental
way to use ABC's rewriting and fraiging routines to do some light-weight
simplification of the AIG that arises from symbolic simulation; the simplified
problem is then solved with ABC's built-in SAT solver or by giving it to a
back-end tool.  It appears that these AIG algorithms may allow GL to scale far
beyond its current capability.

\begin{wrapfigure}[9]{r}[0pt]{2.5in}
\begin{center}
\vspace{-1em}
\begin{tabular}{ll}
\toprule
Independent FADDs     & Total Time    \\
\midrule
AIG Mode              & ? ($>$24 hr)  \\
\satlink/Lingeling     & 18,581 sec  \\
\satlink/Glucose       & 572    sec  \\
ABC/integrated        & 1,446   sec  \\
ABC/\satlink/Lingeling & 442    sec  \\
ABC/\satlink/Glucose   & 121    sec  \\
\bottomrule
\end{tabular}
\end{center}
\end{wrapfigure}

To illustrate, we consider a proof of correctness for a hardware module that
computes four independent single-precision floating point additions in
parallel.  The correctness theorem is stated as a single GL theorem, not broken
into any cases.  This problem is beyond the capacity of our existing BDD or AIG
modes.  Without ABC, only \satlink/Glucose can solve it in a reasonable amount
of time.  When we use ABC to perform some light-weight fraiging and rewriting
first, Glucose can solve the problem almost 5x faster, and the problem also
becomes tractable for other solvers.

\vspace{-.35em}

\section{Conclusions}
\label{sec:conclusion}

\vspace{-.1em}

\haigs and \aignet are two alternatives for representing AIGs that have
complementary strengths and weaknesses.  By keeping the DAG representation
implicit, \haigs provide an especially simple logical story: we can deal
in self-contained Boolean functions rather than a monolithic graph, and we
can reason about AIG operations at the semantic level.  In contrast, by making
the DAG representation explicit, \aignet allows us to write more efficient
algorithms.  Reasoning about these algorithms is more difficult and involves
the kinds of invariants associated with any imperative code, but relationships
like \f{AignetExtension} go a long way toward making this tractable and we have
been able to implement and verify useful \aignet algorithms, e.g., CNF
conversion.

Our work to integrate \aignet into GL allows ACL2 users to make use of
state-of-the-art SAT solvers to handle finite ACL2 goals.  At present these SAT
solvers are simply trusted.  In the future, a verified SAT proof checker might
be integrated into the flow to ensure correct reasoning.

This work is a starting point toward verified implementations of AIG
simplification algorithms like fraiging and rewriting.  We have seen that
(unverified) implementations of these algorithms can significantly reduce the
difficulty of problems that GL gives to the SAT solver, so this is a promising
direction for increasing the scale of theorems that GL can automatically prove.

% Our verified CNF conversion algorithm is a first step towards implementing
% algorithms that rely on incremental SAT.

The source code for our AIG representations, algorithms, and GL connection are
included in the ACL2 Community Books for ACL2 6.1; see
\url{http://acl2-books.googlecode.com/}.  Except for the \aignet library we
rely on features specific to ACL2(h), so please see
\texttt{books/centaur/README.html} for important setup information.  Our \haig
representation and algorithms are found in \texttt{centaur/aig} and the \aignet
library is in \texttt{centaur/aignet}.  The \emph{Centaur Hardware Verification
  Tutorial}, available in \texttt{centaur/tutorial}, shows how to use the new \satlink
mode; see especially \texttt{sat.lsp}.

We thank Bob Boyer, Warren Hunt, and Matt Kaufmann for contributing to the
development of ACL2(h).  We thank Matt Kaufmann for improvements to ACL2 in
support of this work, especially related to abstract stobjs and
non-executability.  We thank Matt Kaufmann, David Rager, Anna Slobadov\'{a},
and Nathan Wetzler for their corrections and helpful feedback on drafts of this
paper.

%BOZO acknowledge Brayton/Mischenko for, e.g., the ABC source code?

%\vspace{-.45em}

\bibliographystyle{eptcs}
\bibliography{paper}
\end{document}